\documentclass[useAMS,usenatbib]{mn2e}

\RequirePackage[dvips]{graphicx}

\def\apjs{\emph{ApJS}}

\title[]{Analytical Approach for the
Determination of the Luminosity Distance in a Flat Universe with
Dark Energy}
\author[T. Wickramasinghe and T. N. Ukwatta]{T. Wickramasinghe$^{1}$ and T. N.
Ukwatta$^{2,3}$\thanks{E-mail:
tilan.ukwatta@gmail.com (AVR)}\\
$^{1}$Department of Physics, The College of New Jersey, Ewing, NJ 08628, USA\\
$^{2}$The George Washington University, Washington, D.C. 20052, USA\\
$^{3}$NASA Goddard Space Flight Center, Greenbelt, MD 20771, USA}

\begin{document}

%\date{Accepted 1988 December 15. Received 1988 December 14; in original form 1988 October 11}

%\pagerange{\pageref{firstpage}--\pageref{lastpage}} \pubyear{2002}

\maketitle

\label{firstpage}

\begin{abstract}
Recent cosmological observations indicate that the present
universe is flat and dark energy dominated. In such a universe,
the calculation of the luminosity distance, $d_L$, involve
repeated numerical calculations. In this paper, it is shown that a
quite efficient approximate analytical expression, having very
small uncertainties, can be obtained for $d_L$. The analytical
calculation is shown to be exceedingly efficient, as compared to
the traditional numerical methods and is potentially useful for
Monte-Carlo simulations involving luminosity distances.
\end{abstract}


\section{Introduction}
The most recent cosmological observations indicate that the
present universe is flat and vacuum dominated~\citep{Komatsu2009}.
In such a vacuum dominated space-time, the distance analysis
requires computer intensive numerical calculations. Even though,
computers today are very fast, efficient analytical calculation of
distance scales would be very useful for various types of Monte
Carlo simulations.

The most fundamental distance scale in the universe is the
luminosity distance, defined by $d_L = \sqrt{L/(4 \pi f)}$, where
$f$ is the observed flux of an astronomical object and $L$ is its
luminosity. Current astronomical observations indicate that the
present density parameter of the universe satisfy $\Omega_\Lambda
+ \Omega_M = 1$ with $\Omega_\Lambda \sim 0.7$. Here
$\Omega_\Lambda$ is the contribution from the vacuum and
$\Omega_M$ is the contribution from all other fields. The distance
calculations in such a vacuum dominated universe involve repeated
numerical calculations and elliptic
functions~\citep{Eisenstein1997}.

In order to simplify the numerical calculations, \cite{Pen1999}
(hereafter Pen99) has developed quite an efficient analytical
recipe. In this paper, we show another analytical method, similar
in many respect to that of Pen99, that can be used to calculate
the distances in a vacuum dominated flat universe.

Our analytical calculation is shown to run faster than that of
Pen99 and has smaller error variations with respect to redshift
($z$) and $\Omega_\Lambda$.

%The paper is organized as follows. In $\S$2, I will develop a
%theory to deduce the luminosity distance from first principles. In
%$\S$3, I will derive necessary tools for the analysis of errors in
%the analytical approach and will show that the errors at z = 1 are
%of the order of 0.05\%, which is decreasing even further as the
%redshift z increases beyond unity, making the analytical algorithm
%exceedingly efficient.

%In this paper we show that the luminosity distance, $d_L$ can be
%obtained to a high degree of accuracy in a relatively simple
%purely analytical method.

Our recipe for calculating the luminosity distance is the
following ($H_0$ is the present Hubble constant and $c$ is the
speed of light):

\begin{equation}\label{eq1}
d_L = \frac{c}{3 H_0} \frac{1+z}{\Omega_\Lambda^{1/6}
(1-\Omega_\Lambda)^{1/3}}[\Psi(x(0, \Omega_\Lambda)) - \Psi(x(z,
\Omega_\Lambda))],
\end{equation}
\begin{equation}\label{eq2}
\Psi(x) = 3 \, x^{1/3} 2^{\,2/3}
\bigg[1-\frac{x^2}{252}+\frac{x^4}{21060}\bigg],
\end{equation}
\begin{equation}\label{eq3}
x=x(z, \Omega_\Lambda) = \ln (\alpha + \sqrt{\alpha^2-1}),
\end{equation}
\begin{equation}\label{eq4}
\alpha=\alpha(z, \Omega_\Lambda) =
1+2\,\frac{\Omega_\Lambda}{1-\Omega_\Lambda}\frac{1}{(1+z)^3}.
\end{equation}

\section[]{Approximation}

We first begin by analyzing how the scale factor, $a(t)$ varies as
a function of time $t$ in a flat universe in which $\Omega_\Lambda
\neq 0$. In this case, $a(t)$ is given by~\citep{Weinberg2008}
\begin{equation}\label{eq:no3}
\dot a^2 = H_{0}^{2} \Omega_{\Lambda} a^{2} + H_{0}^{2} \Omega_{m}
\frac{a^{3}_{0}}{a},
\end{equation}
where $a_0$ is the present value of the scale factor. The above
equation is then immediately integrated into
\begin{equation}\label{eq:no4}
\bigg(\frac{a}{a_0}\bigg)^3 = \frac{1}{2}
\frac{\Omega_m}{\Omega_\Lambda} \bigg[\cosh (3 H_{0} t
\sqrt{\Omega_\Lambda}) -1\bigg].
\end{equation}
The scale factor is directly related to the $z$ as,
\begin{equation}\label{eq:no4b}
\frac{a}{a_0} = \frac{1}{1+z}.
\end{equation}
Let us define $x = 3 H_{0} t \sqrt{\Omega_\Lambda}$ and indicate
its present value by $x_0=x(0, \Omega_\Lambda)$. Then, equations
\ref{eq:no4} and \ref{eq:no4b} give
\begin{equation}\label{eq:no5}
x=x(z, \Omega_\Lambda) = \cosh^{-1}
\bigg[1+2\,\frac{\Omega_\Lambda}{1-\Omega_\Lambda}\frac{1}{(1+z)^3}\bigg]
\end{equation}
If we define $\alpha$ as follows
\begin{equation}\label{eq4}
\alpha=\alpha(z, \Omega_\Lambda) =
1+2\,\frac{\Omega_\Lambda}{1-\Omega_\Lambda}\frac{1}{(1+z)^3}
\end{equation}
and since $\alpha > 1$ we can write $x$ as
\begin{equation}\label{eq3}
x=x(z, \Omega_\Lambda) = \ln (\alpha + \sqrt{\alpha^2-1}).
\end{equation}
We note that $x$ is a monotonically decreasing function beyond
$x(0, 0.7) = 2.42$.

We choose the standard Robertson-Walker
metric~\citep{Weinberg2008} as the metric of the background
space-time. With usual notation, this is

\begin{equation}\label{eq:no6}
ds^2 = c^2 dt^2 - a^2 \bigg[\frac{dr^2}{1-kr^2} + r^2 (d\theta^2 +
\sin^2 \theta \, d\phi^2) \bigg].
\end{equation}

In the above space-time, we can use equation \ref{eq:no3} to
obtain $r$. A straightforward integration for a flat universe ($k
= 0$) yields,

\begin{equation}\label{eq:no7}
r = \frac{c}{a_0 H_0} \frac{1}{3 \Omega_\Lambda^{1/6}
\Omega_M^{1/3}} \int_{x}^{x_0} \frac{dx'}{[\sinh
\frac{x'}{2}]^{2/3}}.
\end{equation}

This integral can be evaluated in terms of hypergeometric
functions and related elliptic integrals. But here we take an
alternate, simple approach by defining a new function,

\begin{equation}\label{eq:no8}
\Psi(x) = \lim_{\delta \rightarrow 0} \int_{\delta}^{x}
\frac{dx'}{[\sinh \frac{x'}{2}]^{2/3}}.
\end{equation}

In the standard model the luminosity distance is defined as $d_L =
a_0 r (1+ z)$. Now we can use equation \ref{eq:no8} to write the
luminosity distance as

\begin{equation}\label{eq:no9}
d_L = \frac{c}{3 H_0} \frac{1+z}{\Omega_\Lambda^{1/6}
\Omega_M^{1/3}}[\Psi(x_0) - \Psi(x)].
\end{equation}

Expanding $\Psi$ in a series expansion to the 4th order, we find
that
\begin{equation}\label{eq:no10}
\Psi(x) = 3 \, x^{1/3} 2^{\,2/3}
\bigg[1-\frac{x^2}{252}+\frac{x^4}{21060}\bigg] + \Psi(0),
\end{equation}
where $\Psi(0) = -2.210$. Now, equation \ref{eq:no9} reduces to
the required expression for the luminosity distance as
\begin{equation}\label{eq:no11}
d_L = \frac{c}{3 H_0} \frac{1+z}{\Omega_\Lambda^{1/6}
(1-\Omega_\Lambda)^{1/3}}[\Psi(x_0) - \Psi(x)].
\end{equation}

\begin{figure}
%\epsscale{1.0}
\includegraphics[width=84mm]{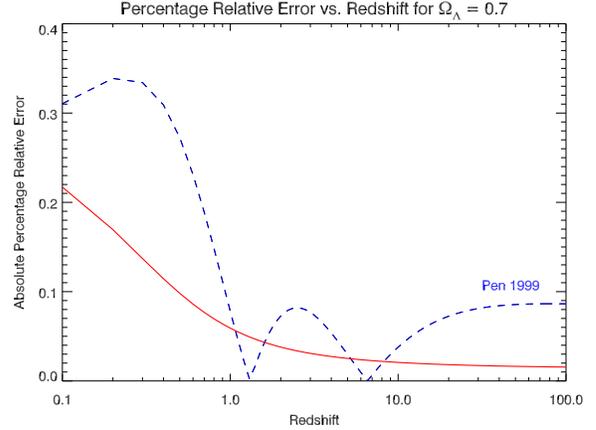}
\caption{The absolute relative percentage error ($\Delta E$) as a
function of the redshift for $\Omega_\Lambda =
0.7$.}\label{WickPenCompare}
\end{figure}

\begin{figure}
%\epsscale{1.0}
\includegraphics[width=84mm]{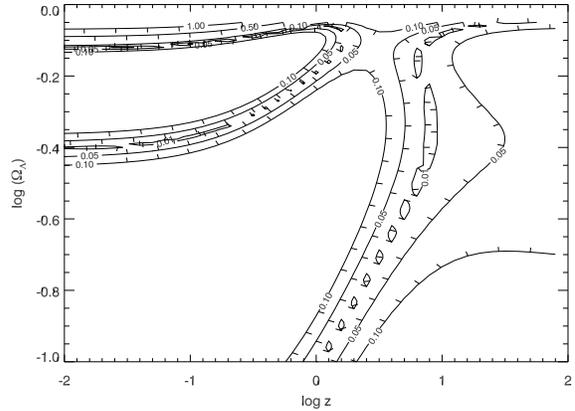}
\caption{Contour plot of absolute relative percentage error
($\Delta E$) for the method of Pen99 with various $z$ and
$\Omega_\Lambda$.}\label{ContourPen}
\end{figure}

\begin{figure}
%\epsscale{1.0}
\includegraphics[width=84mm]{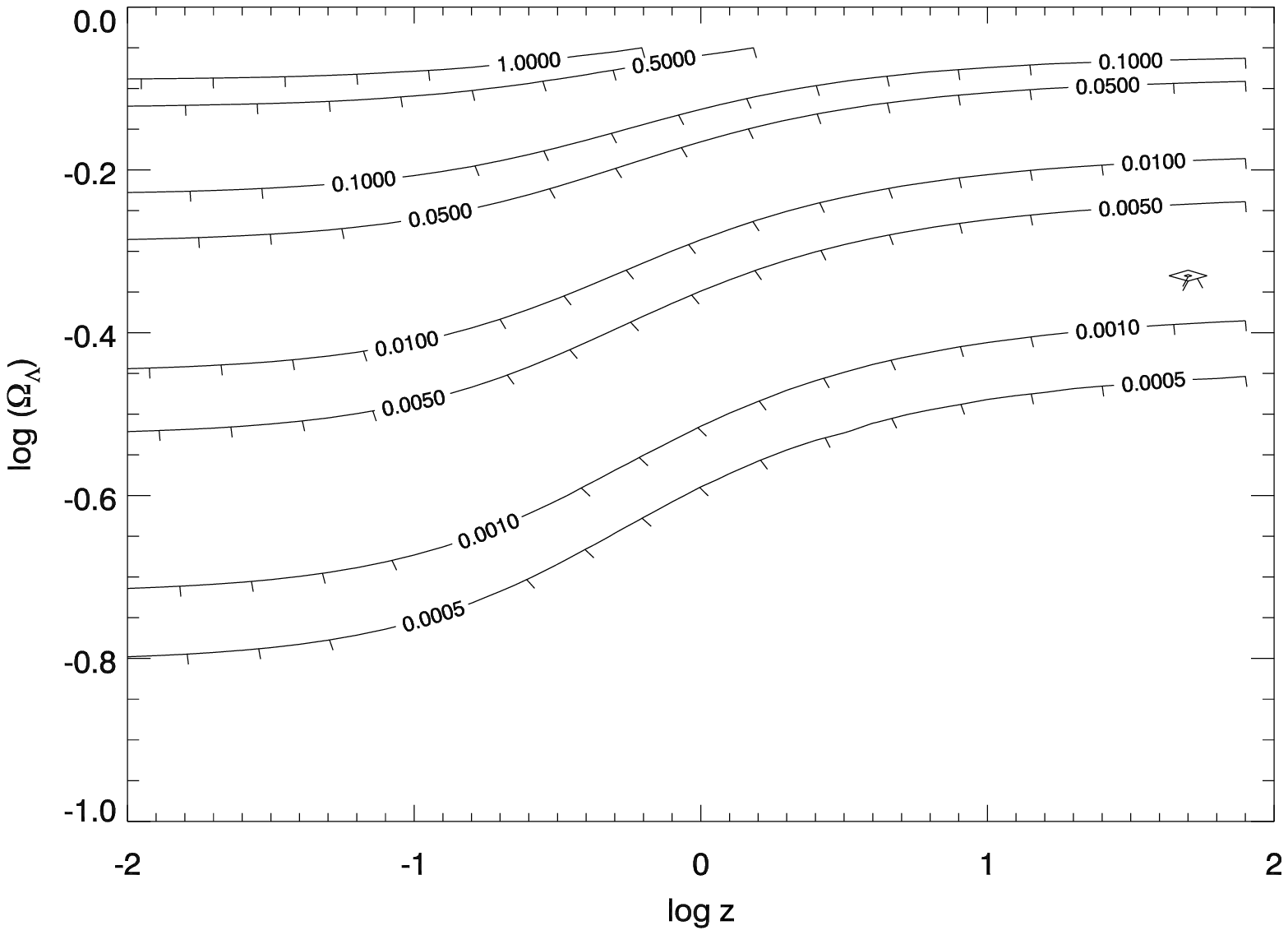}
\caption{Contour plot of absolute relative percentage error
($\Delta E$) for our method with various $z$ and
$\Omega_\Lambda$.}\label{ContourWick}
\end{figure}

\section{Analysis and Conclusion}

In order to compare the method of Pen99 to ours, lets define the
absolute relative percentage error as follows.
\begin{equation}\label{eq:no12}
\Delta E = \frac{|d_{L}^{\tiny \textrm{approx}} - d_{L}^{\tiny
\textrm{num}}|}{d_{L}^{\tiny \textrm{num}}} \times 100
\textrm{\%}.
\end{equation}
Here $d_{L}^{\tiny \textrm{approx}}$ and $d_{L}^{\tiny
\textrm{num}}$ are luminosity distance values calculated from
approximate analytical methods and numerical method respectively.

A comparison of $\Delta E$ for both analytical methods for
$\Omega_\Lambda = 0.7$ is shown in figure~\ref{WickPenCompare}.
Our method has a better absolute relative percentage error value
for $z < 1.0$, $1.6 < z < 5.5$ and $z > 8.0$ compared to that of
Pen99. We note that the error in our method decreases steadily
with redshift approaching $< 0.014$ \% at $z=1100$. In comparison,
for high redshifts, Pen99 error always stays $\sim$ 0.09\% and
does not decrease appreciably.

A contour plot of $\Delta E$ based on the method of Pen99 with
various $z$ and $\Omega_\Lambda$ is shown in
figure~\ref{ContourPen}. Relatively complicated distribution of
variations in the $\Delta E$ can be seen for the parameter space
characterized by $z$ and $\Omega_\Lambda$. However, a contour plot
of $\Delta E$ for our method, which is shown in
Figure~\ref{ContourWick}, shows a smooth behavior over the same
parameter space.

\begin{figure}
%\epsscale{1.0}
\includegraphics[width=80mm]{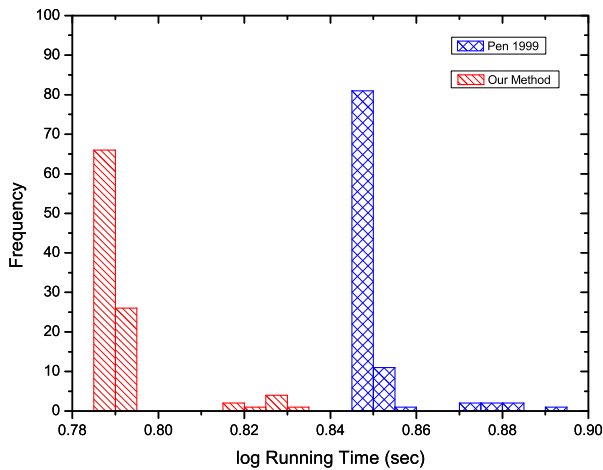}
\caption{Histogram of running times of both Pen99 and our
methods.}\label{HistoPenWick}
\end{figure}

In order to investigate the running time of the two analytical
methods we performed the following test. With $z=1$ and
$\Omega_\Lambda = 0.7$, we calculated the running time for 1
million calculations on a typical personal computer (Intel Core 2
Processor, 2127 MHz, 1 GB RAM, IDL\footnote{Interactive Data
Language
\\
\texttt{http://www.ittvis.com/ProductServices/IDL.aspx}} Version
6.2 running on Windows XP Service Pack 3). Then we repeated the
above process 100 times for the both methods. Histogram of both
running time results are shown in figure~\ref{HistoPenWick}. Our
method is significantly faster than the method of Pen99. In
addition, we performed the same test on the numerical method and
found that our method is more than an order of magnitude faster.
However, we note that the above test is hardware and compiler
dependent and results may vary depending on the hardware and the
compiler used.

With less than 0.1\% error, our analytical method becomes quite
desirable as the most interesting astronomical phenomena happen at
$z > 1$ ($\Omega_\Lambda \sim 0.7$). Furthermore, the analytical
computation is more elegant and faster compared to traditional
numerical computations invoked in connection with calculations of
distances in a vacuum dominated flat universe.

Once we know the luminosity distance, it becomes a simple matter
to evaluate the other distances such as the angular diameter
distance or the proper distance.

%\section*{Acknowledgments}

\end{document}